\documentclass{vgtc}                          

\graphicspath{{figures/}{pictures/}{images/}{./}} 

\usepackage{times}                     

\usepackage{tabu}                      
\usepackage{booktabs}                  
\usepackage{lipsum}                    
\usepackage{mwe}                       

\usepackage{mathptmx}                  

\usepackage{balance}

\onlineid{1159}

\vgtccategory{Research}

\vgtcinsertpkg

\title{Comparing Pass-Through Quality of Mixed Reality Devices: A User Experience Study During Real-World Tasks }

\author{
Francesco Vona\thanks{e-mail: francesco.vona@hshl.de}\\ %
    \scriptsize Hochschule Hamm-Lippstadt %
    \and 
Julia Schorlemmer\thanks{e-mail: julia.schorlemmer@hshl.de}\\ %
     \scriptsize Hochschule Hamm-Lippstadt %
     \and 
Michael Stern\thanks{e-mail: michael.stern@hshl.de}\\ %
    \scriptsize Hochschule Hamm-Lippstadt %
    \and
Navid Ashrafi \thanks{e-mail: navid.ashrafi@hshl.de}\\ %
    \scriptsize Hochschule Hamm-Lippstadt %
    \and
Maurizio Vergari \thanks{e-mail: maurizio.vergari@tu-berlin.de}\\ %
    \scriptsize Technische Universität Berlin %
    \and
Tanja Kojic \thanks{e-mail: tanja.kojic@tu-berlin.de} \\ %
    \scriptsize Technische Universität Berlin %
    \and 
Jan-Niklas Voigt-Antons \thanks{e-mail: jan-niklas.voigt-antons@hshl.de} \\ %
    \scriptsize Hochschule Hamm-Lippstadt %
}

\abstract{
   In extended reality, “pass-through” enables users to view their real-world surroundings via cameras on the headset, displaying live video inside the device. This study compared the pass-through quality of three devices: Apple Vision Pro, Meta Quest 3, and Varjo XR-3. Thirty-one participants performed two tasks—reading a text and solving a puzzle—while using each headset with the pass-through feature activated. Participants then rated their experiences, focusing on workload and cybersickness. Results showed that the Apple Vision Pro outperformed the Meta Quest 3 and Varjo XR-3, receiving the highest ratings for pass-through quality.
} 

\renewcommand\copyrighttext{%
  \footnotesize \textcopyright 2025 IEEE. Personal use of this material is permitted. Permission from IEEE must be obtained for all other uses, in any current or future media, including reprinting/republishing this material for advertising or promotional purposes, creating new collective works, for resale or redistribution to servers or lists, or reuse of any copyrighted component of this work in other works. DOI and link to original publication will be added as soon as they are available.}
  
\newcommand\copyrightnotice{%
   \fbox{\parbox{\dimexpr\columnwidth-\fboxsep-\fboxrule\relax}{\copyrighttext}}
}


\begin{document}

\maketitle
\copyrightnotice
\section{Introduction and Related work}

In June 2023, Apple introduced the Apple Vision Pro, a Mixed Reality (MR) headset with advanced Video See-Through (pass-through) technology. This feature enables seamless transitions along the Reality-Virtuality (RV) continuum \cite{milgram1994}, blending real and virtual worlds through external cameras. Alongside other XR devices like the Varjo XR-3, Pico 4, Meta Quest 3, and Meta Quest Pro, the Apple Vision Pro emphasizes the growing demand for seamless transitions between augmented and virtual realities, with potential applications in education, healthcare, and manufacturing \cite{egger2023apple}. A critical challenge in XR technology remains its ability to support real-world tasks while ensuring user comfort, as traditional VR headsets often isolate users from their physical surroundings. 
This study investigates the pass-through capabilities of three XR devices—Apple Vision Pro, Meta Quest 3, and Varjo XR-3—focusing on user-oriented factors such as cognitive load, cybersickness, and pass-through quality metrics like clarity, depth perception, and resolution.
Pass-through technology has evolved greatly since its debut on the Oculus Quest 2 in 2021, which utilized basic infrared cameras, resulting in low-quality grayscale visuals with distortions \cite{10.1007/s10055-024-00953-w}. Nouri et al. noted such limitations in a study on mixed reality piano teaching applications, emphasizing the need to improve short-distance pass-through \cite{10.1109/MMUL.2022.3232892}. Recent advancements, including NeuralPassthrough \cite{xiao2022}, Passthrough+ \cite{Chaurasia2020}, and techniques addressing VR pass-through without reprojection \cite{Kuo2023}, have enhanced realism and reduced visual discomfort. 
Similarly, Ishihara et al. integrated parallax and latency compensation, improving depth accuracy in video see-through devices \cite{Ishihara2023}.
Applications of pass-through now span education, healthcare, transportation, and entertainment. McGill et al. demonstrated MR headsets’ advantages for spatial awareness in driving tasks despite challenges like motion sickness and social acceptability \cite{McGill2022-gn}. A comparative study found pass-through-enabled VR setups more intuitive and efficient than AR devices for interactive mapping \cite{12}.
However, challenges such as Visually Induced Motion Sickness (VIMS) remain significant \cite{VIMS2020}. Methodological improvements, like controlled trial durations and breaks, have been suggested to mitigate discomfort \cite{VIMS2020}. Finally, broader surveys \cite{Koulieris2019, Itoh2021} underscore the importance of advancing display technologies and balancing innovation with user comfort and safety.
\vspace{-0.3cm}
\begin{figure}[ht]
 \centering 
\includegraphics[width=\columnwidth, height = 4.5cm]{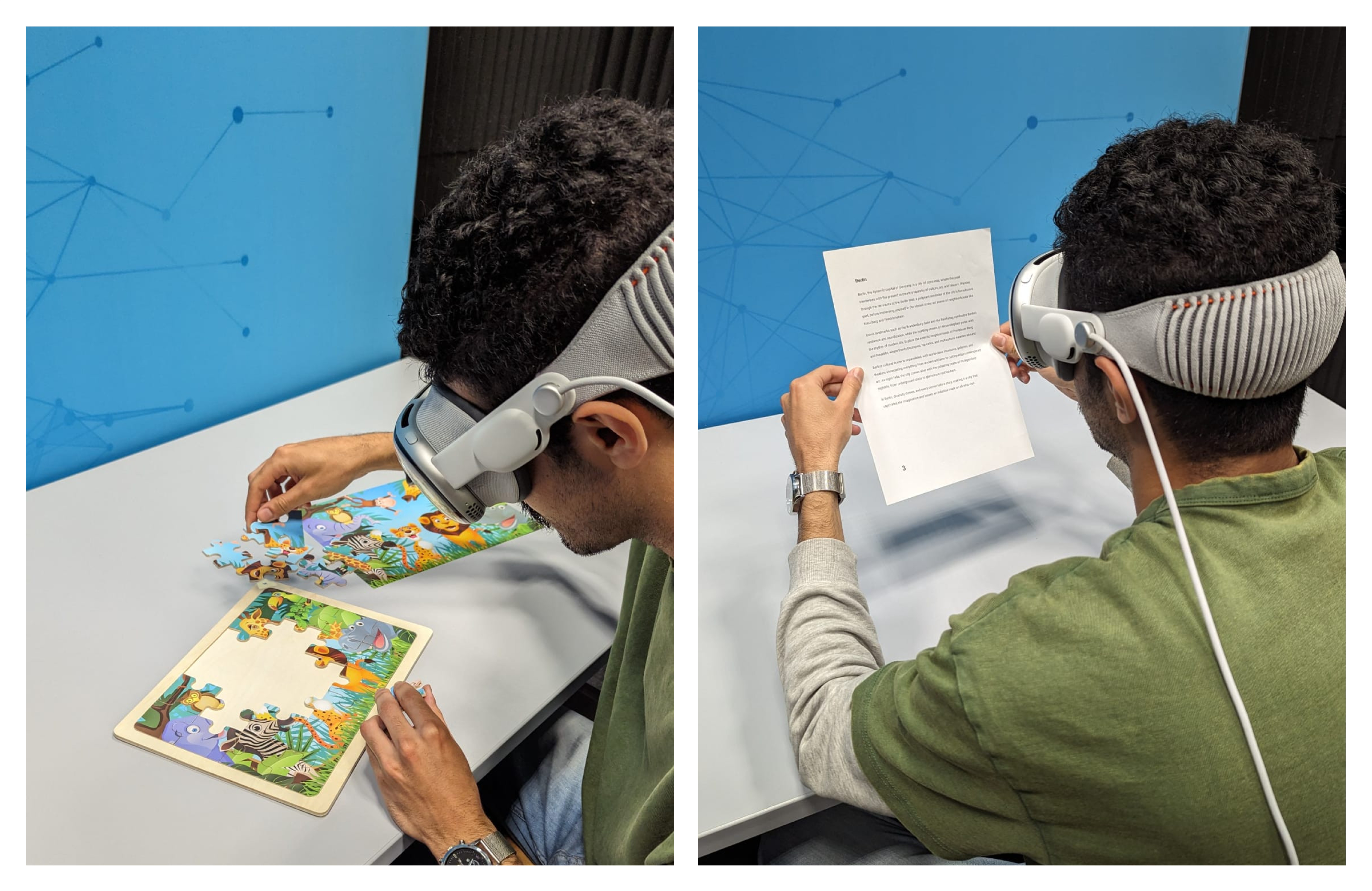}
\vspace{-0.7cm}
\small \caption{Participant engaged in the puzzle task (left) and the reading task (right) while wearing the AVP headset}
  \label{fig:exp_task}
\vspace{-0.4cm}
\end{figure}

\begin{figure*}[t]
 \centering
 \includegraphics[alt={Box-plots showing the distribution of total NASA TLX scores and CSQ-VR scores across the different conditions.}, width=\textwidth, height=4cm]{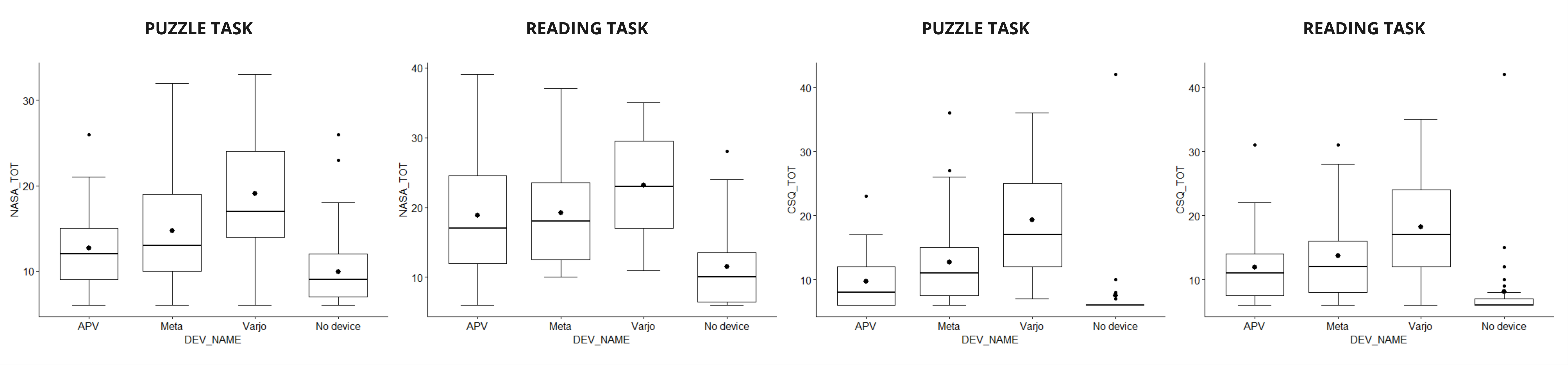}
  \vspace{-0.7cm}
 \small \caption{Box-plots showing the distribution of total NASA TLX scores and CSQ-VR scores across the different conditions.}
 \label{fig:anova}
 \vspace{-0.41cm}
\end{figure*}

\section{Methods}
A within-subjects study was conducted with 31 participants (18 male, 13 female, M = 30.06 years, SD = 6.57). Participants were asked to complete two tasks under four different conditions. The tasks involved solving puzzles and reading texts aloud, while the conditions included three headsets: \textit{Apple Vision Pro}, \textit{Meta Quest 3}, \textit{Varjo XR-3}, and a fourth baseline condition with no device/headset. 
Conditions were randomized to minimize order effects. Performance was assessed using: i) NASA TLX: To evaluate cognitive load \cite{HART1988139} (\textit{NASATOT}); ii) Cybersickness Questionnaire (\textit{CSQ-VR}): To measure discomfort across nausea (\textit{CSQ-NAU}), vestibular (\textit{CSQ-VES}), and oculomotor (\textit{CSQ-OCU}) dimensions \cite{virtualworlds2010002}; iii) Pass-through Quality Metrics: Custom metrics evaluated clarity, resolution, color accuracy, depth perception, environmental awareness, and distortion (\textit{CLA, RES, CA, DP, ENV, DIS}).
The study was conducted in a controlled laboratory setting at the local university campus after receving the approval from the university's ethics board. Each participant was tested individually, with sessions lasting approximately one hour. During the experiment, participants were seated at a table and instructed to either read a text aloud or solve a puzzle in each condition. After completing each task, participants filled out the NASA TLX, the CSQ-VR, and the pass-through quality questions, repeating this for each of the four conditions. When a headset was used, an experimenter set up the device, activated the pass-through, and cleaned the headset after each use.

\section{Results}
\balance
\textbf{Task Load.} Apple Vision Pro recorded the lowest task load for both reading (M = 18.84, SD = 8.66) and puzzle-solving (M = 12.71, SD = 4.32), compared to Meta Quest 3 (M = 19.26, SD = 8.17; M = 14.71, SD = 6.22) and Varjo XR-3 (M = 23.16, SD = 7.31; M = 19.10, SD = 7.19). Repeated-measures ANOVA revealed significant differences in task load across conditions ($F(3, 90) = 28.12, p < 0.05, \eta^2 = 0.24)$ (Figure~\ref{fig:anova}). \textbf{Cybersickness. }The Apple Vision Pro exhibited minimal cybersickness (M = 9.68, SD = 4.26), followed by the Meta Quest 3 (M = 12.68, SD = 7.02), with the Varjo XR-3 experiencing the highest discomfort (M = 19.35, SD = 8.60). Repeated-measures ANOVA revealed significant differences in task load across conditions $(F(3, 90) = 16.11, p < 0.05, \eta^2 = 0.21)$ (Figure \ref{fig:anova}).
\textbf{Pass-Through Quality.} Apple Vision Pro registered the highest score in all metrics, including clarity (M = 5.39), resolution (M = 5.55), and environmental awareness (M = 5.58). The Meta Quest 3 performed moderately well, while the Varjo XR-3 recorded the lowest scores. Repeated-measures ANOVA for pass-through dimensions revealed significant differences as well. 

\section{Discussion and Conclusion}
The Apple Vision Pro’s performance underscores the importance of high-quality displays, accurate motion tracking, and ergonomic design in enhancing user experiences. Its low task load and minimal cybersickness make it suitable for long-term applications, a critical factor for adoption in fields like healthcare and education \cite{egger2023apple}.
The Meta Quest 3 demonstrated a balanced performance suitable for moderate use, though higher task load and cybersickness suggest a need for refinement. In contrast, the Varjo XR-3’s higher cognitive demands and discomfort highlight the challenges of optimizing pass-through quality, particularly in-depth perception and distortion reduction.
These findings align with prior research emphasizing the role of display resolution and latency reduction in improving XR experiences \cite{Fernandes2016, Rebenitsch2016}. This study highlights the Apple Vision Pro’s better quality in pass-through technology, offering unmatched clarity, resolution, and user comfort. The Meta Quest 3 provides an adequate alternative but requires cognitive load and cybersickness improvements. Varjo XR-3’s limitations ask for further refinement to compete in practical applications. Future research should include diverse devices (e.g., Meta Quest Pro) and explore the long-term effects of pass-through use. Other points on our agenda include testing with a larger sample size and different tasks. 

\bibliographystyle{abbrv-doi}

\bibliography{main}
\end{document}